\documentclass[aip,pop,preprint]{revtex4-1}
\usepackage[utf8]{inputenc}
\usepackage{amsmath,amssymb,siunitx}
\usepackage{graphicx,float,dcolumn,bm}
\usepackage{hyperref,comment}
\usepackage{natbib}
\begin{document}

\title[The physics of PLX as modeled with FLASH, OSIRIS, and HELIOS]{Exploring the physics of the Plasma Liner Experiment: A Multi-dimensional study with FLASH, OSIRIS, and HELIOS}

\author{E.~C.~Hansen}
\affiliation{University of Rochester, Rochester, New York 14627, USA}
\author{P.~Farmakis}
\affiliation{University of Rochester, Rochester, New York 14627, USA}
\author{D.~Michta}
\affiliation{University of Rochester, Rochester, New York 14627, USA}
\author{C.~Ren}
\affiliation{University of Rochester, Rochester, New York 14627, USA}
\author{H.~Wen}
\affiliation{University of Rochester, Rochester, New York 14627, USA}
\author{S.~Langendorf}
\affiliation{Los Alamos National Laboratory, Los Alamos, New Mexico 87545, USA}
\author{F.~Chu}
\affiliation{Los Alamos National Laboratory, Los Alamos, New Mexico 87545, USA}
\author{P.~Tzeferacos}
\affiliation{University of Rochester, Rochester, New York 14627, USA}

\date{\today}

\begin{abstract}
The Plasma Liner Experiment (PLX) at Los Alamos National Laboratory (LANL) is a platform that seeks to achieve fusion via the Plasma-Jet-Driven Magneto-Inertial Fusion (PJMIF) concept.
The experiment utilizes a constellation of plasma guns to generate fusion-relevant conditions and consists of three main phases: (1) target formation, in which up to four plasma guns shoot magnetized hydrogen or deuterium-tritium jets to form a quasi-spherical target, (2) liner formation, in which a 36 guns fire high-atomic-number (e.g., xenon) jets to form a liner shell, and (3) target compression, in which the formed liner implodes the pre-formed target.
Each phase of the PLX operates in different plasma regimes, with different physics at play, thus each phase must be simulated separately with appropriate codes.
In this study we highlight 1-, 2-, and 3-D simulation results of all three phases using the \emph{FLASH}, \emph{OSIRIS}, and \emph{HELIOS} codes.
Some of the key physical processes involved include shock dynamics, kinetic effects, anisotropic thermal conduction, resistive magnetic diffusion, radiation transport, and magnetized jet dynamics.
The simulations show that the PLX can form a preheated ($\sim$40 eV), magnetized (electron Hall parameter $>$1) target plasma, and a quasi-collisional liner shell, which can subsequently compress the target to fusion-relevant conditions, reaching temperatures in excess of  1 keV.
\end{abstract}

\maketitle

\section{\label{sec:intro}Introduction}
The pursuit of practical fusion energy remains a grand scientific and engineering challenge, with several promising approaches under active investigation \cite{ikeda2007progress, buttery2019diii, zylstra2022burning, abu2024achievement}.
Among them, Plasma-Jet-Driven Magneto-Inertial Fusion (PJMIF) represents a hybrid method that seeks to combine the advantages of both magnetic and inertial confinement fusion \cite{Hsu2012, Thio2019}.
This novel concept employs high-velocity plasma jets to form a spherically symmetric liner, which compresses a magnetized fuel target to fusion conditions.
The PJMIF approach has the potential to achieve high energy gain at relatively low driver energies, making it an attractive alternative to conventional fusion concepts.

The Plasma Liner Experiment (PLX) at Los Alamos National Laboratory (LANL) serves as a critical testbed for investigating the feasibility of PJMIF \cite{merritt2014experimental, langendorf2018experimental, lajoie2024formation}.
The heavy, fast-moving plasma liner used to drive the implosion of the plasma target is formed from the merging and spherical convergence of an array of 36 supersonic plasma jets traveling at $\sim 50$ km/s \cite{yates2020experimental}.
Recent advancements have improved our understanding of the PLX properties, such as liner uniformity and jet merging dynamics, and have guided the development of novel, optimized jet configurations and injection parameters to enhance liner performance \cite{lajoie2023multi, lajoie2024formation}.
A compatible magnetized plasma target that takes advantage of the high implosion speed of a spherically imploding plasma liner is injected into the vessel by merging $4-6$ magnetized plasma jets, whose magnetic field strength is $\sim$ 1 kG at the gun nozzles.
The collision of these plasma jets is expected to induce turbulence in the plasma target, resulting in tangled fields, which can in turn provide very long conduction lengths between the core and the liner surface thereby reducing thermal conduction losses from the fuel plasma to the colder liner \cite{ryutov2009adiabatic, hsu2019magnetized, chu2023characterization}.

The conditions in PJMIF implosions span a wide range of plasma parameter space and physical regimes, ranging from kinetic to fluid collisionality, optically thin to thick radiation transport, and weak to moderate magnetization.
Some of the pertinent physical processes include cold, high-velocity jets merging at oblique angles during liner formation, target formation of a preheated magnetized plasma, anisotropic magnetized thermal conduction, and ultimately fusion thermonuclear burn and charged particle transport \cite{chu2023experimental}.
Due to the wide range of relevant physics, previous studies in PJMIF have often employed analytical or semi-analytical approaches to predict performance. \cite{Langendorf2017, Hsu2019}
Integrated simulation efforts have been attempted and have addressed many first-order questions \cite{Awe2011, Davis2012, Thoma2013, Schillo2020, Thompson2020}, but have often been limited in the breadth of the physical effects they could include.
In this work, an attempt is made to rigorously simulate as many aspects of the concept as feasible with leading modern computational capabilities, and give improved insight into the detailed performance of PJMIF conceptual designs.

The structure of the paper is as follows:
in Section~\ref{sec:methods} we describe the three codes used in this work: \emph{FLASH}, \emph{OSIRIS} and \emph{HELIOS}.
Then, the paper is divided by the three phases of the PLX: target formation (Section~\ref{sec:targ}), liner formation (Section~\ref{sec:liner}), and target compression (Section~\ref{sec:compression}).
Lastly, we summarize our conclusions in Section~\ref{sec:conc}.

\section{\label{sec:methods}Numerical Methods}
For the radiation magneto-hydrodynamic (MHD) modeling of the various stages of the PLX platform we employ the \emph{FLASH} code \cite{Fryxell2000}.
\emph{FLASH} is a publicly-available, parallel, multi-physics, adaptive mesh refinement (AMR), finite-volume Eulerian hydrodynamics and MHD code, developed at the University of Rochester by the Flash Center for Computational Science (for more information on the \emph{FLASH} code, visit: \protect\url{https://flash.rochester.edu}).
\emph{FLASH} scales well to over 100,000 processors in modern high-performance computing architectures, and uses a variety of parallelization techniques like domain decomposition, mesh replication, and threading, to optimally utilize hardware resources.
The \emph{FLASH} code has a world-wide user base of more than 5,000 scientists, and more than 1,300 papers have been published using the code to model problems in plasma astrophysics, combustion, computational fluid dynamics, high energy density physics (HEDP), and fusion energy research.

Over the past decade and under the auspices of the U.S. Department of Energy (DOE) National Nuclear Security Administration (NNSA), the Flash Center has added in \emph{FLASH} extensive HEDP and extended-MHD capabilities~\cite{Tzeferacos2015} that make it an ideal tool for the multi-physics modeling of the PLX platform.
These include state-of-the art hydrodynamic and MHD shock-capturing solvers,~\cite{Lee2013} extended to a three-temperature formalism,~\cite{Tzeferacos2015} anisotropic thermal conduction that utilizes high-fidelity magnetized heat transport coefficients,~\cite{JiHeld2013} electron-ion heat exchange, multi-group radiation diffusion, tabulated multi-material equations of state (EOS) and opacities, and numerous synthetic diagnostics \cite{Tzeferacos2017}.
\emph{FLASH}'s newest algorithmic developments also include a complete generalized Ohm's law that incorporates all extended-MHD terms of the Braginskii formulation. \cite{Braginskii1965}
The new extended MHD capabilities are integrated with state-of-the-art transport coefficients,~\cite{Davies2021}  developed with support from the U.S. DOE Advanced Research Projects Agency–Energy (ARPA-E) BETHE program.
The \emph{FLASH} code and its capabilities for magnetized HEDP have been verified through several benchmarks and code-to-code comparisons,~\cite{Fatenejad2013, Orban2013, Orban2022, Sauppe2023} as well as through direct application to numerous plasma physics experiments,~\cite{Meinecke2014, Meinecke2015, Li2016, Tzeferacos2018, Rigby2018, White2019, Chen2020,  Bott2021, Meinecke2022} leading to innovative science and publications in high-impact journals.

For the kinetic modeling of target and liner plasma-jet merging processes we employ the \emph{OSIRIS} code.\cite{Fonseca02,hemker00}
\emph{OSIRIS} is a state-of-the-art, fully explicit, multi-dimensional (1D, 2D slab and $r$-$z$, and 3D), fully parallelized, fully relativistic, particle-in-cell (PIC) code.
\emph{OSIRIS} is jointly developed by Instituto Superior Técnico Lisboa and the University of California, Los Angeles (for more information on the \emph{OSIRIS} code, visit: \protect\url{https://osiris-code.github.io}).
\emph{OSIRIS} is highly optimized on a single core, and it scales well to over 1.5 million cores on the Sequoia supercomputer at Lawrence Livermore National Laboratory, maintaining 97\% and 78\% efficiency on weak and strong scaling, respectively.
It has been run on several of the U.S. DOE and National Science Foundation (NSF) leadership-class computing facilities for over 15 years.
On Blue Waters it has achieved 2.2 Pflops on a full-machine benchmark using more than 10 trillion particles.
The \emph{OSIRIS} code is used world-wide by more than 25 user groups to model problems in laser-plasma interaction, wake field acceleration, ion acceleration, plasma astrophysics such as shock formation, and fusion energy research.

\emph{OSIRIS} includes higher order particle shapes, which limits self-heating effects related to aliasing, and includes energy- and momentum-conserving algorithms.
The parallelization is done either using domain decomposition with MPI only or by using a hybrid approach where MPI is used across nodes and OpenMP parallelization is used within a node; both approaches support dynamic load balancing at the MPI level.\cite{Miller2021}
The dynamic load balancing feature can greatly speed up the plasma-jet collision simulations because the compute time for each time step is roughly proportional to the number of particles in the MPI nodes.
The spatial distribution of the number of particles changes drastically as the simulations progress.
There is a relativistic two-body Coulomb collision model \cite{nanbu1997,takizuka1977,peano2009} as well as a hybrid algorithm \cite{Fiuza2011} which pushes particles throughout all regions, but in a high-density region (i.e., collisional), MHD field equations are used to solve for the fields.
This is useful in simulations of plasmas with a large density range, where some parts of the system are collisionless while others are collisional.

Results are also compared with \emph{HELIOS}, a lightweight 1D Lagrangian fluid code that includes physics packages useful for HEDP such as tabulated EOS, flux-limited radiation diffusion, and thermal conduction. \cite{MacFarlane2006}
Of particular interest in this study, the 1D Lagrangian formulation exactly preserves material interfaces (e.g., between the liner and the target) with no numerical diffusion or mixing between species.
In a realistic case, mixing may occur due to various physical mechanisms, such as the Rayleigh-Taylor instability, but the comparison of 1D Lagrangian-to-Eulerian codes allows for a convenient way to gauge the impact of numerical diffusion in the Eulerian solve, as a function of grid resolution, separate from physical processes.

\section{\label{sec:targ}Target Formation}
To model the targer formation phase of the PLX, we ran a 2D axisymmetric \emph{FLASH} simulation of two counter-propagating jets to demonstrate how the target plasma is formed from the jet collision.
The code solves the three-temperature resistive MHD equations with anisotropic thermal conduction, electron-ion heat exchange, and multi-group radiation (with six energy groups).
For the jets, the new implementations of thermal conductivity\cite{JiHeld2013} and magnetic resistivity\cite{Davies2021} coefficients are used, coupled with \emph{PROPACEOS} tables for the frequency-dependent opacity.
The domain extends axially from -137 cm to 137 cm (i.e., the diameter of the PLX chamber), and spans 68.5 cm radially.
The simulation employs adaptive mesh refinement to achieve a maximum effective resolution of nearly 32 cells across the initial outer jet radius.

The jets are modeled as fully-ionized hydrogen with an ideal EOS closure ($\gamma = 5/3$).
The jets are initialized with an inner radius of 1.5 cm (i.e., the plasma gun wire radius) and an outer radius of 4.25 cm.
Between these radii, the jets are given a mass density of $2\times10^{-8}$ g cm\textsuperscript{-3}, a velocity of $8\times10^6$ cm s\textsuperscript{-1}, electron and ion temperatures of 1 eV, and a radiation temperature of 0.1 eV.
These parameters are enforced as inflow conditions at the lower and upper $z$ boundaries of the computational domain, which are also treated as solid reflecting walls from $r = 0$ to 1.5 cm and beyond 4.25 cm.
The solid-wall regions of the domain boundaries do not allow any hydrodynamic motion, but they allow thermal, radiation, and magnetic diffusion to pass through them.
A radially-varying azimuthal magnetic field is imposed for the entire radial extent at these boundaries. That magnetic field has a peak value of 2,500 G at 1.5 cm, with a linear profile from $r = 0$ cm to 1.5 cm, a $1/r$ profile from 1.5 cm to 4.25 cm, and a value of zero beyond 4.25 cm.
This magnetic field profile is also imposed at every time step on the first row of cells inside the domain, which helps reduce the imbalance of hydromagnetic pressure with the inflow conditions for better control of the jet velocity.

Since Eulerian codes like \emph{FLASH} do not simulate a true vacuum, the PLX chamber or ``ambient'' region is initialized with a low density of $2\times10^{-11}$ g cm\textsuperscript{-3}, where thermal conduction and radiation absorption and emission are turned off.
The electron and ion temperatures are initialized to 1 eV and radiation temperature to 0.1 eV everywhere inside the domain.
The difference in density between the vacuum and jet material creates a pressure imbalance that increases the forward jet velocity and also causes the jets to expand radially into the vacuum.
There are also $\mathbf{J} \times \mathbf{B}$ forces from the azimuthal magnetic field that pinch the jet towards $r = 0$ during its propagation.
Once the pinched jet material reaches the symmetry axis, it has no choice but to move axially, which leads to a streamlining of the jet head, observed as an overall acceleration of the jet; this is a well-documented phenomenon sometimes referred to as the nose-cone effect (see \citet{Hansen2015} and references therein).
The combined effects of the enforced jet inflow velocity, pressure imbalance, and $\mathbf{J} \times \mathbf{B}$ pinch forces result in an average jet velocity (from $t=0$ s to collision) of about 1.25 $10^7$ cm s\textsuperscript{-1}, which is higher than the reported experimental values \cite{Hsu2015} but not necessarily inconsistent.
Previous experiments measured the jet velocities at positions near the gun and slightly downstream and reported a possible range of 0.3-1 $10^7$ cm s\textsuperscript{-1}.
Given the aforementioned acceleration mechanisms, an average value of 1.25 $10^7$ cm s\textsuperscript{-1} over the entire axial extent is plausible.

\begin{figure}
  \includegraphics[width=0.7\linewidth]{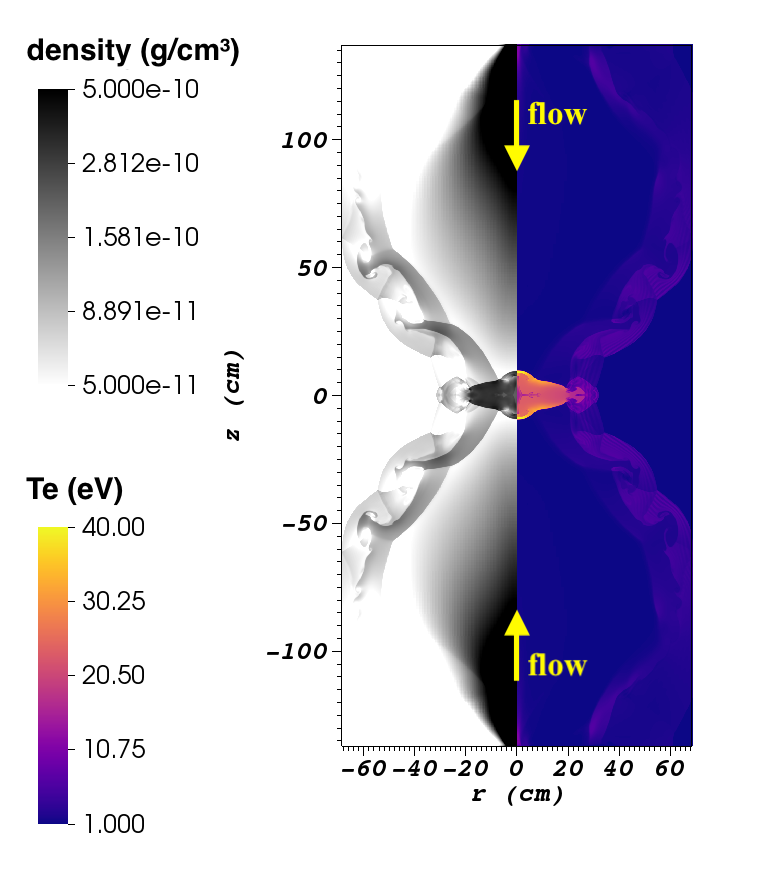}
  \caption{Mass density (g cm\textsuperscript{-3} and electron temperature (eV) of two colliding jets that have formed a target plasma from a 2D axisymmetric \emph{FLASH} simulation.
  This target reaches a volume-averaged density of $1.67\times10^{-10}$ g cm\textsuperscript{-3} and peak preheat temperatures of around 40 eV.}
  \label{fig:jets_tele}
\end{figure}

Fig.~\ref{fig:jets_tele} shows a density and electron temperature map of the jets after they have collided and formed the target plasma.
Note that some amount of jet material is left behind or expands outwards away from the target-forming region.
This effect can be mitigated by adding more jets; the experimental platform currently employs four target-forming jets, but may use more in the future.
Also, some of the jet material will be swept up by the liner-forming jets during the subsequent liner-forming process.
As the jets collide, material is forced outwards radially, creating an elongated, pancake-like target.
This effect can also be mitigated by adding more jets to the formation process, which would result in a more spherical target.
By the end of the simulation, the target reaches a volume-averaged density of approximately $1.67\times10^{-10}$ g cm\textsuperscript{-3}.

The temperature map of Fig.~\ref{fig:jets_tele} shows that the hottest plasma forms a shell around the newly formed target.
These outermost regions of the target plasma are similar to accretion shocks where jet material piles up and flows into the expanding target.
Since the magnetic field is azimuthal and the hot shell has a magnetic resistivity that is lower than that of the jets, the magnetic field has increased difficulty leaving the shell region.
The magnetized shell can assist in thermally insulating the target via reduced thermal conductivity, which is included in the 
anisotropic thermal conduction implementation in \emph{FLASH}.
Some amount of preheat prior to the liner compression of the target is crucial to obtaining the desired plasma properties - we observe peak temperatures of $\sim$ 40 eV in the jet collision simulation.

\begin{figure}
  \includegraphics[width=0.6\linewidth]{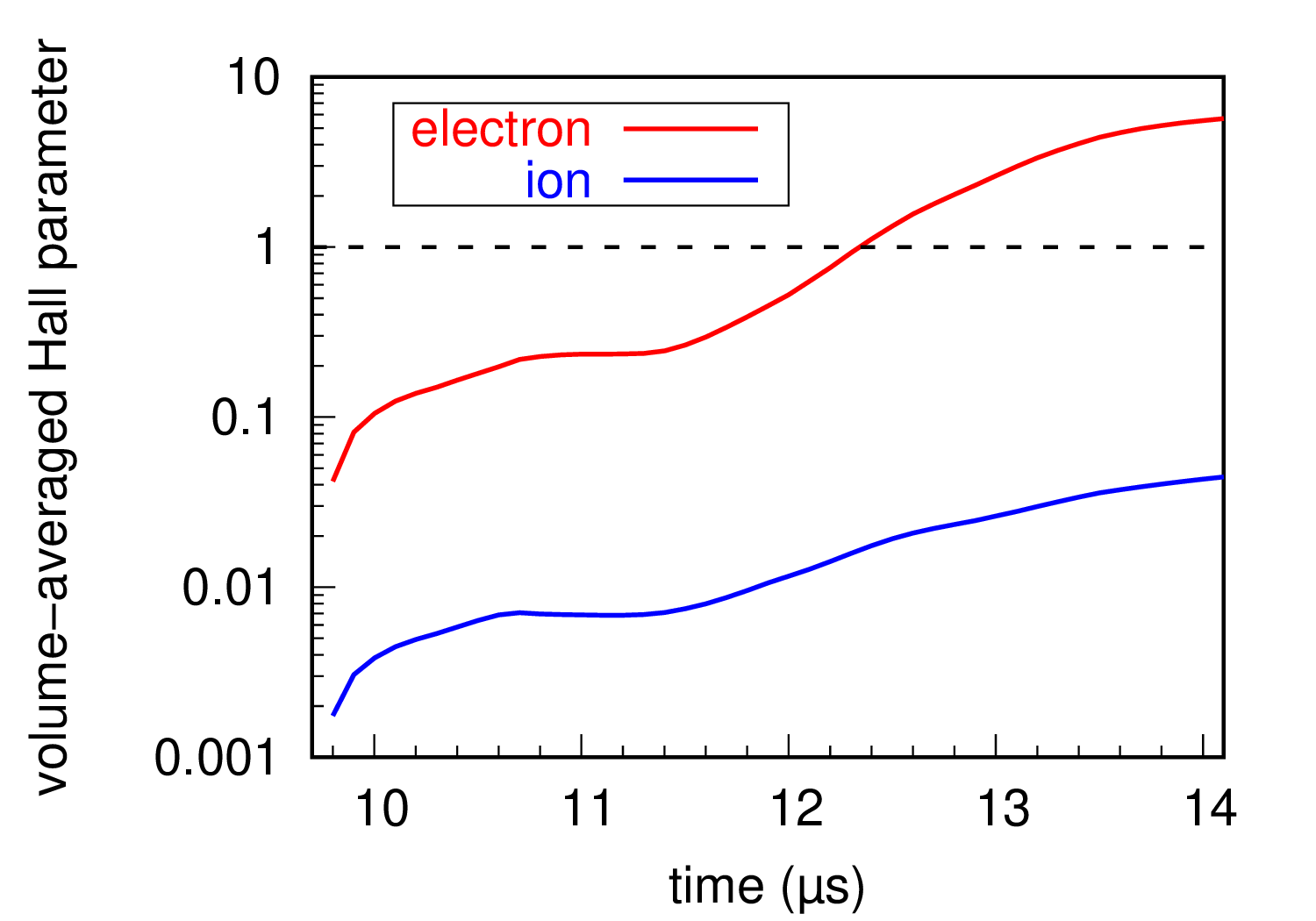}
  \caption{Volume-averaged electron (red) and ion (blue) Hall parameters from the 2D jet collision simulation as a function of time.
The values are averaged in a region where the target plasma is forming from $r = 0$ to $r = 25$ cm and $z = -15$ to $z = 15$ cm.
Charged particles are considered magnetized when their Hall parameter is above unity (denoted by the black dashed line), which in this case eventually occurs for electrons but not for ions.}
  \label{fig:jets_chi}
\end{figure}

The postprocessing analysis included volume-averaging several quantities over a region about the center of the domain from $r = 0$ to $r = 25$ cm and $z = -15$ to $z = 15$ cm.
In this region, the target attains a volume-averaged number density of approximately $1\times10^{14}$ cm\textsuperscript{-3}, which would increase experimentally with more jets.
We also expect the final volume-averaged magnetic field strength, 24.4 G, to also increase with the number of jets.
Fig.~\ref{fig:jets_chi} shows the volume-averaged Hall parameters for electrons ($\langle \chi_e \rangle$) and for ions ($\langle \chi_i \rangle$) as a function of time.
The target plasma reaches $\langle \chi_e \rangle > 5.0$, which is consistent with previous \emph{FLASH} and \emph{OSIRIS} simulations of target formation \cite{Wen2022} that reported electron Hall parameters greater than unity.
Also, the attained average plasma $\beta$ (i.e., the ratio of thermal-to-magnetic pressure) was slightly greater than 100.
This indicates that the magnetic field is strong enough to assist in thermally insulating the target plasma while not being dynamically significant, which is one of the primary goals of the PLX target formation process.

The results from this jet collision simulation are used to inform the initial conditions for 3D experiment-scale target compression simulations presented later in this work (Section~\ref{subsec:flash3d_compression}).

\section{\label{sec:liner}Liner Formation}
In the full experiment-scale PLX, the liner shell is formed using 36 plasma guns, but some of the key physics can be studied by simulating the merging of just two jets.
The primary goal of our liner formation simulations was to understand which conditions lead to fluid-like, collisional regimes and which lead to more kinetic-like, collisionless regimes.
This jet-merging process in the context of PJMIF was recently studied by \citet{Cagas2023}, and their results were consistent with some early findings from the PLX.
Fluid codes (such as \emph{FLASH}) will always produce shocks when the two jets collide, but these shocks can be mitigated through the collisionless interpenetration of the jets, a process that requires kinetic modeling.

Here we have chosen to simulate the merging of two liner jets with both a particle-in-cell code (\emph{OSIRIS}) and a fluid code (\emph{FLASH}) to make direct comparisons that help quantify the collisionality of the liner formation phase.

\subsection{\label{subsec:osiris_liner}\emph{OSIRIS} Simulations}
\begin{figure}
  \includegraphics[width=\linewidth]{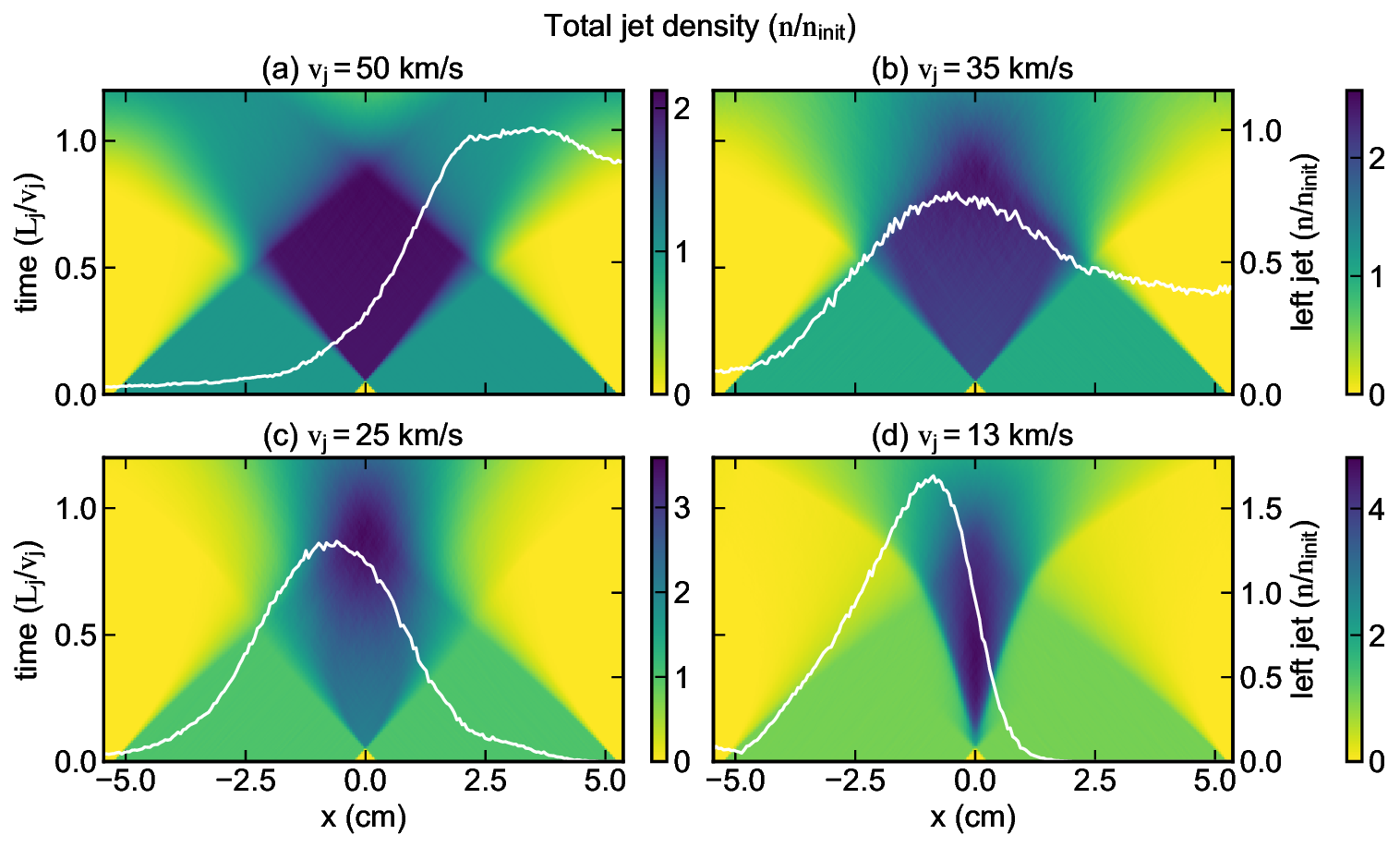}
  \caption{Density evolution of two Xe liner plasma jets modeled using 1D \emph{OSIRIS}.
  The horizontal axis is position (cm), left vertical axis is time in units of jet crossing times, and the total density (normalized to the initial value) is represented by the color map.
  The white lines represent the density profile of the left jet at the end of the simulation, normalized to the initial value and plotted with the right vertical axis; the location of the peak indicates how much the jets have interpenetrated.
  More interpenetration was observed at higher jet velocities.}
  \label{fig:liner_osiris}
\end{figure}

A series of 1D \emph{OSIRIS} simulations were conducted to quantify the interpenetration during the liner-liner jet merging process.
A transition from a collisional to collisionless regime was observed as the initial jet velocity was increased.
In these simulations, two singly-charged xenon plasma jets with density $n_\mathrm{Xe}=4.13\times10^{16}$ $\mathrm{cm}^{-3}$ were initialized with equal electron and ion temperatures $T_\mathrm{e}=T_\mathrm{Xe}=1.5$ eV.
The two jets counter-propagated with velocity $v_\mathrm{j}$, where $v_\mathrm{j}$ is the controlled parameter.
Fig.~\ref{fig:liner_osiris} shows the evolution of the plasma jet density, where the white lines correspond to the density profile of the left jet at the end of the simulation.
At high jet velocities (e.g., $v_\mathrm{j}=5\times10^6$ cm $\mathrm{s}^{-1}$), the merging of the two jets was collisionless; the left jet was found to fully penetrate through the right jet, as indicated by the white line in Fig.~\ref{fig:liner_osiris}(a)  (i.e., the final left jet density profile peaks on the right side of the domain).
At low jet velocities (e.g., $v_\mathrm{j}=1.3\times10^6$ cm $\mathrm{s}^{-1}$), the merging of the two jets was collisional; the left jet was mostly stopped by the right jet with minimal interpenetration as shown in Fig.~\ref{fig:liner_osiris}(d).
At intermediate velocities, the two jets can partially interpenetrate and form a broad density profile that peaks near the center, which is desirable for the PLX design.
These results are consistent with the theoretical estimates of the ion mean free path, which is close to the liner jet size (5 cm) at $v_\mathrm{j}=2.7\times10^6$ cm $\mathrm{s}^{-1}$. 

\subsection{\label{subsec:flash_liner}\emph{FLASH} Simulations}
In this section, 1D liner-liner jet merging simulations conducted with \emph{FLASH} are presented and compared with those from \emph{OSIRIS}. 
This comparison aims to better delineate the transition from collisional to collisionless regimes and to identify any additional effects from radiative losses. 
The simulation setup for the two counter-propagating xenon jets is consistent with the previous subsection, aside from the introduction of a low-density pseudo-vacuum, with a density four orders of magnitude lower than the jet density. 
Both the jets and the pseudo-vacuum are modeled using an ideal gamma-law EOS, with $\gamma=5/3$.  
Thermal conduction and electron-ion heat exchange are described with Spitzer models. \cite{Spitzer1962}  

To incorporate radiation effects, we use \emph{FLASH's} multi-group diffusion radiation transport model. 
The opacities for the jets and the pseudo-vacuum were tabulated using the \emph{PROPACEOS} code. 
The initial radiation temperature was set to be in equilibrium between ions and electrons, at $T_{rad}=1.5$ eV. 
Simulations were conducted both with and without the inclusion of radiation transport to assess the impact of the latter. 

\begin{figure}
  \includegraphics[width=\linewidth]{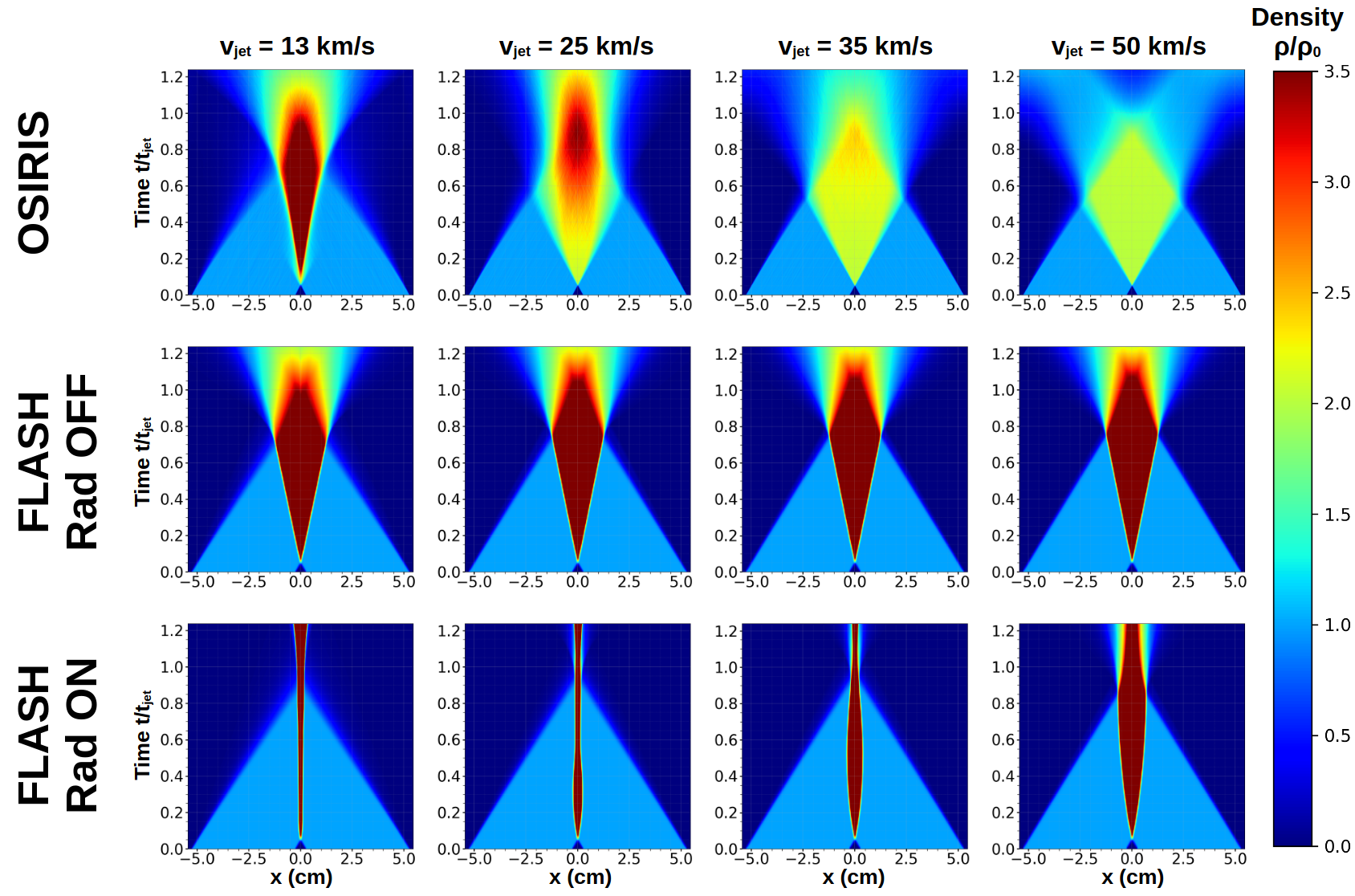}
  \caption{Total density evolution in the 1D liner jet merging simulations using \emph{OSIRIS}, \emph{FLASH} without radiation transport, and \emph{FLASH} with radiation transport, for four jet velocities.
  \emph{FLASH} simulations without radiation transport align with \emph{OSIRIS} in the low-velocity (non-interpenetrating) regime.
  The inclusion of radiation transport in \emph{FLASH} results in a more pronounced collapse of the liner.}
  \label{fig:liner_flash}
\end{figure}

Fig.~\ref{fig:liner_flash} shows the evolution of total density in simulations using \emph{OSIRIS}, \emph{FLASH} without radiation transport, and \emph{FLASH} with radiation transport, at four different jet velocities. 
In the collisional regime, observed at lower jet velocities ($v_j = 1.3\times10^6$ and $2.5\times10^6$ cm $\mathrm{s}^{-1}$), \emph{FLASH} simulations without radiation transport show qualitative similarities with the \emph{OSIRIS} results.
However, the behavior diverges at higher velocities, with \emph{OSIRIS} jets interpenetrating while the \emph{FLASH} jets compressing to a higher density. 
Including radiation transport in \emph{FLASH} modifies the dynamics of the interaction: the addition of radiative losses allows for more compression, leading to further collapse of the liner. This indicates that while the particle-in-cell modeling can more accurately capture the regime transition from collisional to collisionless, radiation losses may also be important during the liner formation. 

\section{\label{sec:compression}Target Compression}
The final phase of the PLX is target compression, during which fusion-relevant conditions may be attained.
We studied several aspects of the target compression process: kinetic effects of a heavy liner colliding with a lighter target plasma, how liner perturbations affect implosion dynamics, and the fusion-relevant performance of an idealized reactor-scale compression.
In all of the simulations that follow, assumptions were made about a pre-formed target plasma and a pre-formed liner plasma shell to study target compression independently from the other PLX phases.

\subsection{\label{subsec:1d_compression}1D Code-to-code Comparisons}
\begin{figure}
  \includegraphics[width=\linewidth]{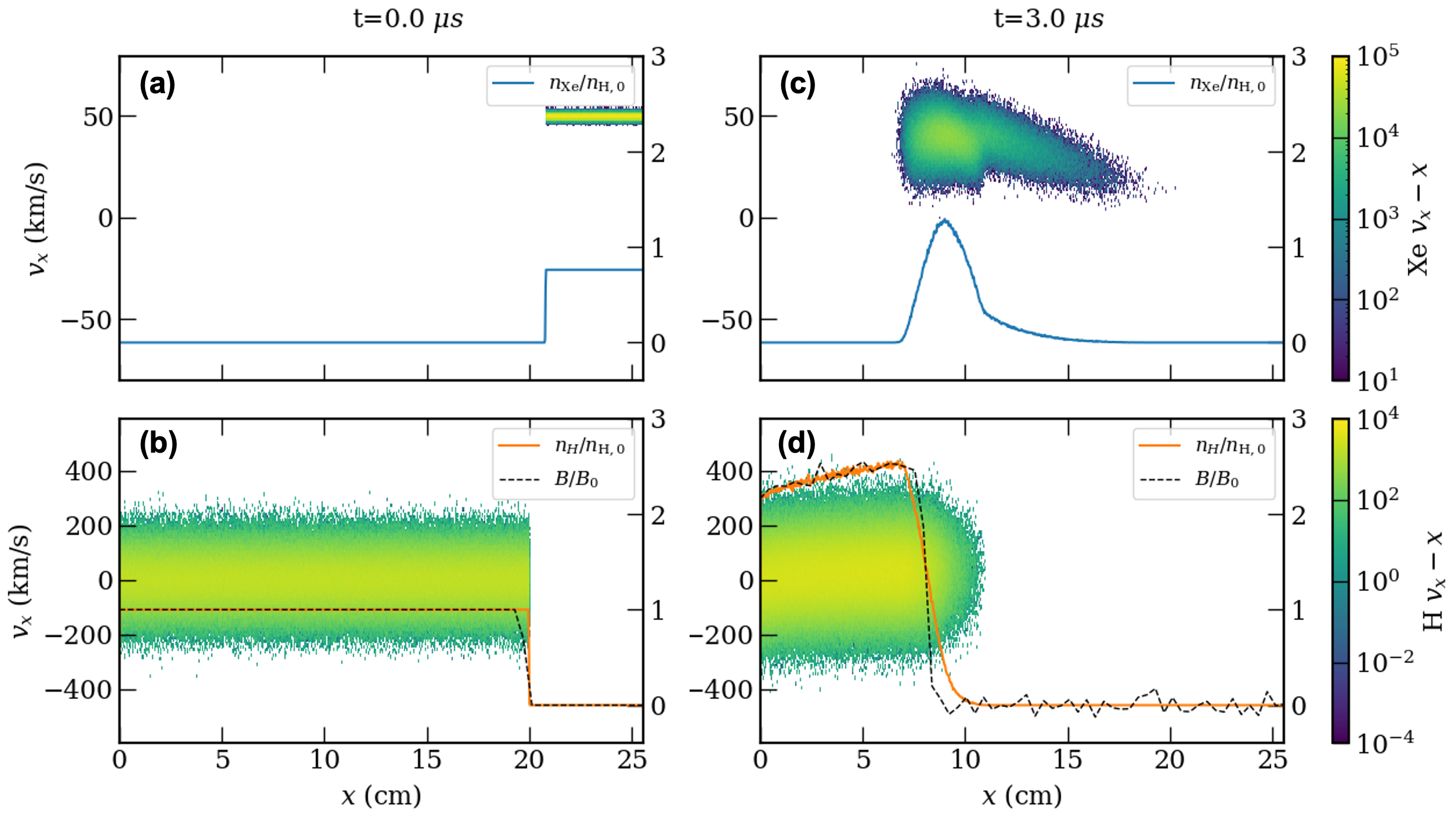}
  \caption{The $v_x$--$x$ phase space (plotted with left vertical axes) of the liner (upper row) and the fuel (bottom row) at $t=0$ (left column) and $t=3.7$ $\mu$s (right column).
  The color bars represents particle number density in arbitrary units, to illustrate where in phase space the liner and fuel plasmas exist. 
  The blue and orange solid lines correspond to the densities of the liner ($n_\mathrm{Xe}$) and fuel ($n_\mathrm{H}$) plasmas, respectively.
  The black dashed line corresponds to the amplitude of the external magnetic field ($B$).
  The densities were normalized to the initial value of the fuel, the external magnetic field was normalized to its initial value (50 T), and they are both plotted with the right vertical axes.}
  \label{fig:os_compression}
\end{figure}

The kinetic process of target compression was investigated with a 1D \emph{OSIRIS} simulation in Cartesian geometry.
The initial condition of the PIC simulation is illustrated in Figs.~\ref{fig:os_compression}(a) and \ref{fig:os_compression}(b).
A 5 cm singly-charged ($Z=1$) xenon liner propagated from right to left with velocity $v_x=5 \times10^6\,\mathrm{cm}\,\mathrm{s}^{-1}$ and collided with a 20-cm, stationary, fully-ionized hydrogen fuel that was magnetized by a 50 T magnetic field.
Reflecting boundary conditions were used for both fields and particles.
The initial densities for the liner and fuel plasma were $4.1\times 10^{16}$ $\mathrm{cm}^{-3}$ and $5.4\times 10^{16}$ $\mathrm{cm}^{-3}$, respectively.
The simulation grid size was on the same order as the Debye length of the fuel with temperature $T_\mathrm{fuel}=50$ eV, which consequently did not resolve the cold liner ($T_\mathrm{liner}=1.5$ eV).
To reduce the numerical heating of the liner, a high-order current-smoothing and field-smoothing scheme \cite{Shang1999} was used.
The total energy increased by about 7\% at the peak compression of the fuel, which is much less than the change in fuel thermal energy.
At $t=0$, the majority of the energy was stored in the form of the liner kinetic energy.
During the liner-fuel compression process, the liner kinetic energy was converted into thermal energy of the liner and fuel.
Both the liner and fuel plasma were heated as evidenced by the broadening of the phase spaces along the $v_x$ axis shown in Figs.~\ref{fig:os_compression}(c) and \ref{fig:os_compression}(d).
As shown in Fig.~\ref{fig:os_compression}(c), the magnetic field was also compressed by the liner, keeping the $B/n_\mathrm{H}$ ratio nearly constant within the fuel region.

\begin{figure}
  \includegraphics[width=\linewidth]{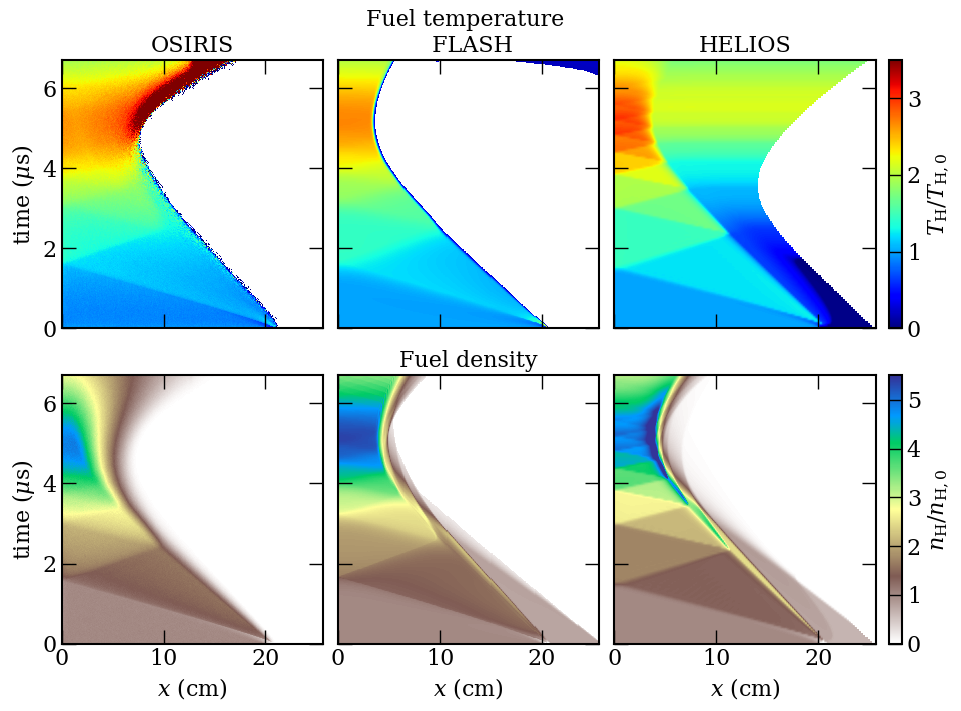}
  \caption{The evolution of the temperature (upper row) and density (bottom row) of the fuel obtained from \emph{OSIRIS} (left column), \emph{FLASH} (middle column), and \emph{HELIOS} (right column) simulations.
  All quantities are normalized to their respective initial values.
  All three codes yielded quantitatively consistent results. }
  \label{fig:compare_3codes}
\end{figure}

The evolution of the fuel density and temperature obtained from the PIC simulation qualitatively agrees with the MHD simulations as illustrated in Fig.~\ref{fig:compare_3codes}.
All three codes predicted peak compression to occur at about $t=4.5$ $\mu s$.
The maximum temperature and density achieved in the \emph{OSIRIS} simulation were slightly lower than those achieved in the \emph{FLASH} and \emph{HELIOS} simulations, which is likely due to mixing of the liner and fuel plasmas.
This mixing also resulted in a shallower density gradient as shown in Fig.~\ref{fig:compare_3codes}.

In the \emph{OSIRIS} simulation, a low-pass filter was applied to the electromagnetic field in the region $x > 18$ cm after $t=3.8$ $\mu s$ to suppress a numerical instability observed in that region.
Since the plasma density in that region is very low ($ < 0.01$ $n_\mathrm{H,0}$) after $t=3.0$ $\mu s$, we do not anticipate the additional low-pass filter to influence the physics of the liner-fuel compression which happened in the region $x < 10$ cm.
Any meaningful influence should travel at the Alfv\`en speed from $x > 18$ cm to $x < 10$ cm, which would take more than 1 $\mu s$.
As shown in Fig.~\ref{fig:compare_3codes}(a), the high temperature in the liner-fuel mixing region after the peak compression is likely an artifact due to an insufficient number of particles in that region.

\begin{figure}
  \includegraphics[width=0.6\linewidth]{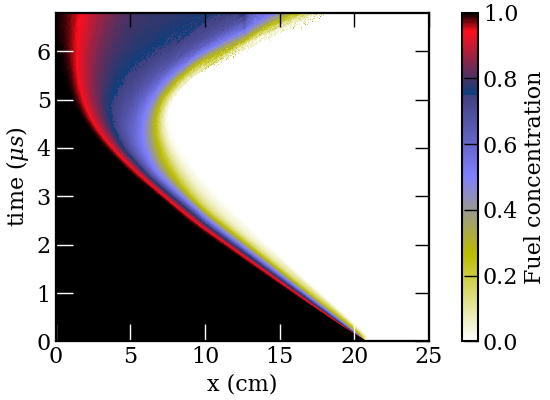}
  \caption{The evolution of the spatially resolved fuel concentration ($n_\mathrm{H}/(n_\mathrm{H}+n_\mathrm{Xe})$) obtained from the \emph{OSIRIS} simulation.
  The fuel-liner mixing is limited in the central part of the fuel region.}
  \label{fig:fuel_fraction}
\end{figure}

Fig.~\ref{fig:fuel_fraction} shows the evolution of the spatially resolved fuel concentration which is defined as $n_\mathrm{H}/(n_\mathrm{H}+n_\mathrm{Xe})$.
We observe evidence of mixing, which helps explain the slightly lower temperatures in the \emph{OSIRIS} results.

\subsection{\label{subsec:flash3d_compression}3D \emph{FLASH} Simulations}
High-resolution 3D simulations would have been too computationally expensive for \emph{OSIRIS}, and \emph{HELIOS} only operates in 1D.
We must therefore turn to the \emph{FLASH} code for high-fidelity 3D simulations of the target compression phase.
Our first goal was to study the effects of liner perturbations on implosion dynamics, since the liner formation process is not likely to produce a perfectly homogeneous shell.

In all three simulations of this subsection, we used results from the 2D \emph{FLASH} target formation simulation (see Section~\ref{sec:targ}) to inform the initial conditions of the target plasma.
Since the previous jet collision simulation used two jets, it was reasonable to multiply the target density by a factor of two to approximate a target formed from four jets, which is the number of target-forming plasma guns employed by the current PLX design.
Therefore, the hydrogen target was initialized with a density of approximately $3.35\times10^{-10}$ g cm\textsuperscript{-3}.
The initial target temperature was set to 40 eV, a level of preheat observed in the target formation simulation.
The target is assumed to initially be perfectly spherical with a radius of 25 cm, resolved with a minimum cell width of 0.125 cm.

Inside the target, we add pseudo-random perturbations $\mathbf{\epsilon}$ to a constant vector potential $\mathbf{A}$, and then $\mathbf{\nabla} \times (\mathbf{A} + \mathbf{\epsilon})$ gives the initial randomized magnetic field.
Each component of $\mathbf{A}$ uses a different pseudo-random number $\alpha_i$, ranging from -1 to 1, for each computational cell.
By using $A_i = 3.5$ G cm for each component of $\mathbf{A}$ and $\epsilon_i = 3.43 \alpha_i$ G cm, the resulting initial magnetic field has a volume-averaged value of roughly 36.1 G, which, like the density, is consistent with an increase from the attained value in the 2D jet collision simulation.
Furthermore, the initial volume-averaged electron Hall parameter is greater than unity and $\langle \beta \rangle$ is greater than 100, which is again consistent with the 2D jet collision simulation.
The field everywhere outside of the target plasma is initialized to zero.

\begin{figure}
  \includegraphics[width=0.8\linewidth]{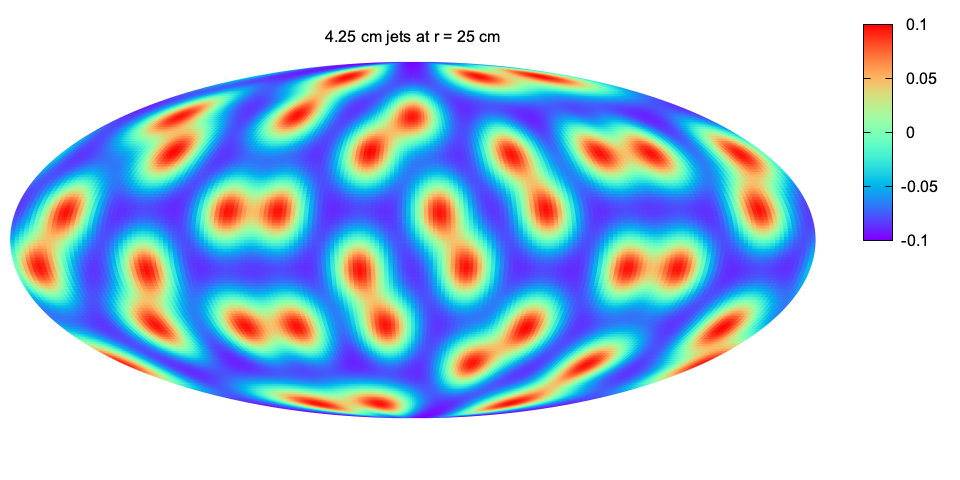}
  \caption{A spherical slice of the mathematical form (dimensionless) of the liner perturbations taken at a radius of 25 cm (the inner liner radius), projected to a 2D map.
  Each spot on the surface is from a Gaussian centered at the experimental angular position of a plasma gun with a spot size that has increased from the initial 4.25 cm jet radius.}
  \label{fig:linepert}
\end{figure}

The Xe liner is initialized with a density of roughly $2.99\times10^{-7}$ g cm\textsuperscript{-3}, temperature of 1.5 eV, velocity of $6\times10^6$ cm $\mathrm{s}^{-1}$, and thickness of 5 cm.
For the simulation cases with liner perturbations, we imposed Gaussian density and velocity perturbations throughout the thickness of the liner at the angular positions of the experimental plasma guns.
An example of this kind of perturbation is shown in Fig.~\ref{fig:linepert}, which displays the mathematical form (dimensionless) of the perturbation as a spherical slice at a radius of 25 cm (the inner liner radius), projected to a 2D map.
The purpose of Fig.~\ref{fig:linepert} is to illustrate that the regions of increased density and velocity are indeed centered at the angular positions of the experimental plasma guns, and that each Gaussian has an overall spot size larger than 4.25 cm (the initial jet radius at the chamber wall).
These perturbations are normalized to have precise control of their amplitude. Three simulations were executed with 0\%, 5\%, and 10\% perturbations.

Outside of the liner is a pseudo-vacuum, which we model as an extension of the Xe liner but with a density and velocity that decrease linearly to $2.99\times10^{-9}$ g cm\textsuperscript{-3} and 0 cm $\textrm{s}^-1$, respectively.
The initial conditions for these simulations are summarized in Table~\ref{tab:3dparams}.

\begin{table}[H]
\centering
\begin{ruledtabular}
\begin{tabular}{ccc}
Parameter & Experiment-scale & Reactor-scale \\
\hline
min. cell width (cm) & 0.12500 & 0.03125 \\
\hline
target radius (cm) & 25 & 4.0 \\
\hline
target $\rho$ (g cm\textsuperscript{-3})  & $3.35\times10^{-10}$ & $2.24\times10^{-5}$ \\
\hline
target $T_e$, $T_i$ (eV) & 40.0 & 108 \\
\hline
target $\langle B \rangle$ (G) & 36.1 & $6.29\times10^{4}$ \\
\hline
target $\langle \chi_e \rangle$ & 22.8 & 9.93 \\
\hline
target $\langle \beta \rangle$ & 427 & 9.82 \\
\hline
liner thickness (cm) & 5 & 1 \\
\hline
liner $\rho$ (g cm\textsuperscript{-3})  & $2.99\times10^{-7}$ & $2.40\times10^{-2}$ \\
\hline
liner $T_e$, $T_i$ (eV) & 1.5 & 1.5 \\
\hline
liner velocity (km/s) & 60 & 80 \\
\hline
liner perturbations & 0\%, 5\%, 10\% & 0\% \\
\hline
vacuum $\rho$ (g cm\textsuperscript{-3})  & $2.99\times10^{-9}$ & $2.40\times10^{-6}$ \\
\end{tabular}
\end{ruledtabular}
\caption{Initial conditions used for the 3D \emph{FLASH} simulations of the experiment-scale and reactor-scale target compression.}
\label{tab:3dparams}
\end{table}

\begin{figure}
  \includegraphics[width=\linewidth]{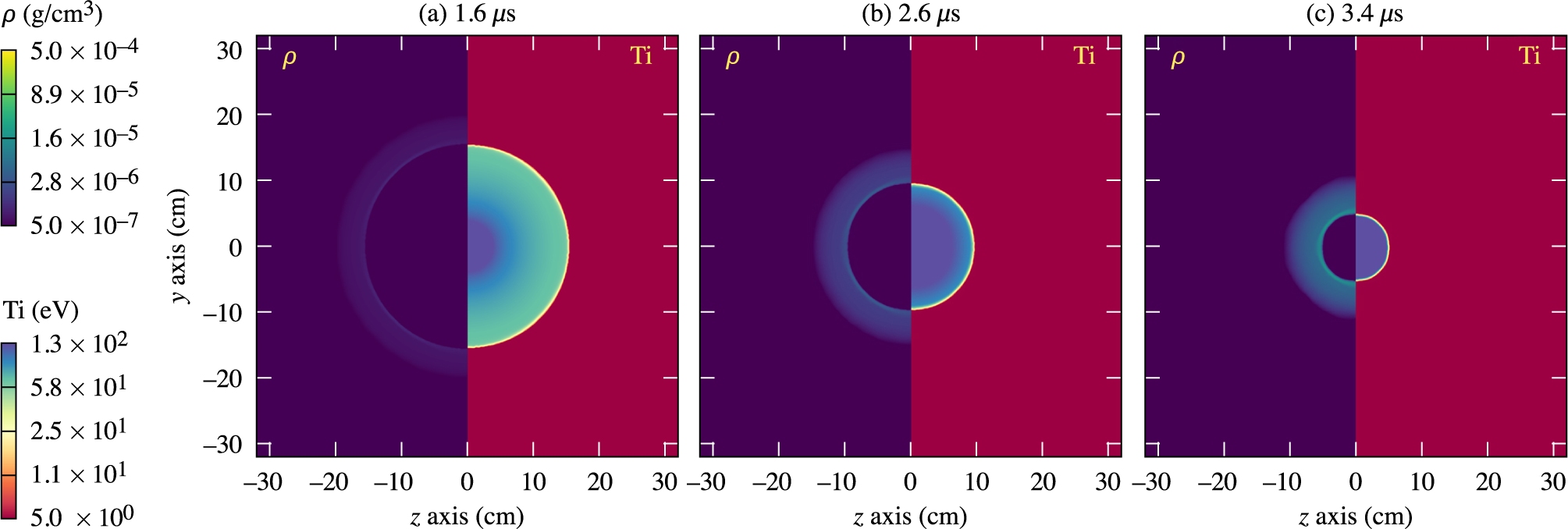}
  \caption{A time series of slices at the $y$-$z$ mid-plane showing mass density in g cm\textsuperscript{-3} (left halves) and ion temperature in eV (right halves) for the 0\% perturbation case.}
  \label{fig:dens_tion_0pct}
\end{figure}

We first look at the 0\% perturbation case in detail to understand the dynamics and relevant physics.
Fig.~\ref{fig:dens_tion_0pct} shows a time series of the implosion, viewed as 2D slices of mass density (g cm\textsuperscript{-3}) and ion temperature (eV).
At the early time of 1.6 $\mu$s, the target density has more than tripled from its initial value, and a temperature gradient has been established by the competing effects of compressional heating and thermal conduction.
The central temperature has increased due to compression, but the target plasma's contact with the colder liner plasma has resulted in some thermal losses.
Magnetization helps reduce thermal losses to some extent, and ion temperatures overall increase.
Target density and ion temperature both increase for the duration of the simulation; volume-averaged values reach 500 times the initial value and $>$150 eV, respectively.
Localized peak values can be higher and the platform can be further optimized to improve performance on a larger reactor-scale platform with more plasma guns.
Without any liner perturbation, the implosion remains spherical.

\begin{figure}
  \includegraphics[width=\linewidth]{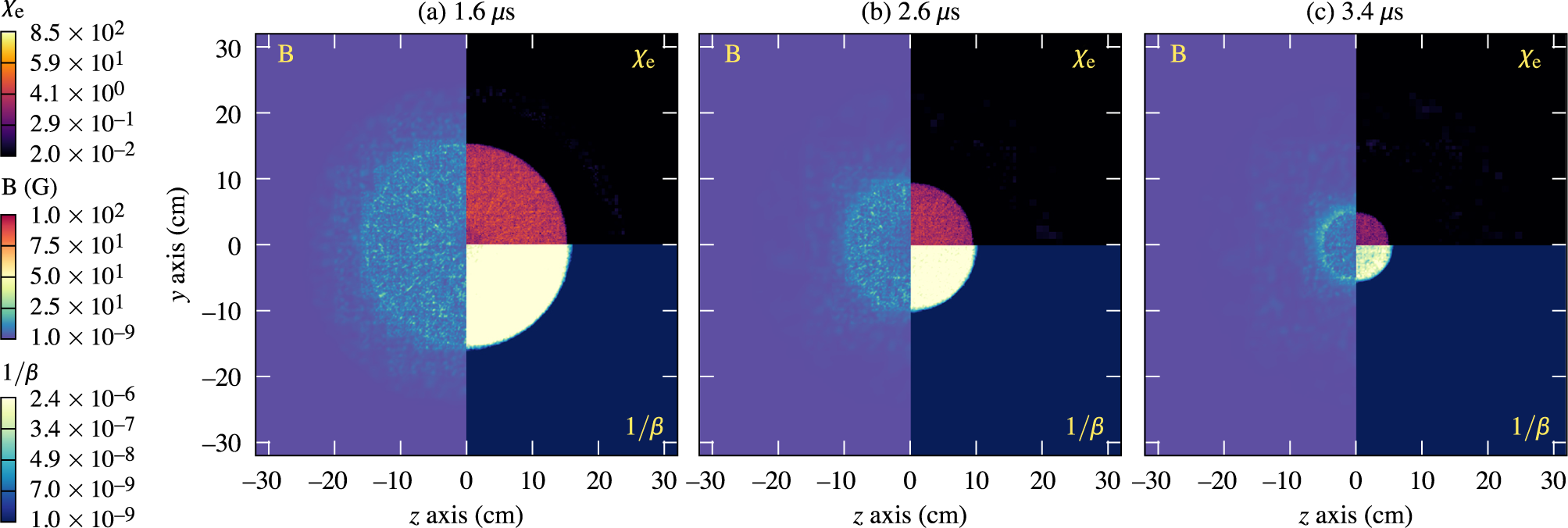}
  \caption{A time series of slices at the $y$-$z$ mid-plane showing magnetic field magnitude in G (left halves), electron Hall parameter $\chi_e$ (upper-right), and inverse plasma beta $1/\beta$ (lower-right) for the 0\% perturbation case.}
  \label{fig:mag_0pct}
\end{figure}

Fig.~\ref{fig:mag_0pct} shows the same time series as the previous figure but now for magnetic field relevant parameters: magnetic field magnitude (G), electron Hall parameter $\chi_e$, and inverse plasma beta $1/\beta$ (the ratio of magnetic-to-thermal pressure).
There are a few key points of note: 1) the randomization of the initial magnetic field results in peaks and valleys that make the plots appear noisy; 2) $\chi_e$ is overall less than unity as some B-field has been lost; and 3) $1/\beta$ remains very small during the simulation, i.e., the magnetic field is never dynamically important.
Although $\chi_e$ has peak values $>$100, these are highly localized to where the magnetic field is also peaked, thus much of the overall benefit of magnetized thermal insulation is lost.
In fact, the volume-averaged $\chi_e$ inside the target drops below unity at about 0.3 $\mu$s.
We can see an enhancement of B-field inside the liner at 3.4 $\mu$s, which suggests that the field has diffused out of the target plasma in part because the cold liner has a relatively high magnetic resistivity.

These results suggest that future PLX experiments should  guard against magnetic field loss due to resistive diffusion.
In addition to simply using stronger initial fields, there are other ways this effect may be mitigated.
More target preheat would decrease the target resistivity and help keep the field more ``frozen-in'' to the target plasma, allowing for more magnetic field compression.
Keeping the liner hotter may also be beneficial, and could be achieved by using a material with a lower atomic number than Xe.
A lower-$Z$ material would inherently have a lower resistivity, but would also radiate less, maintain a relatively smaller resistivity. These effects must be balanced against the benefits of the cold-liner performance as an effective piston for for compression. There is some room for optimization in this parameter space, but such a task is beyond the scope of this work.

\begin{figure}
  \includegraphics[width=\linewidth]{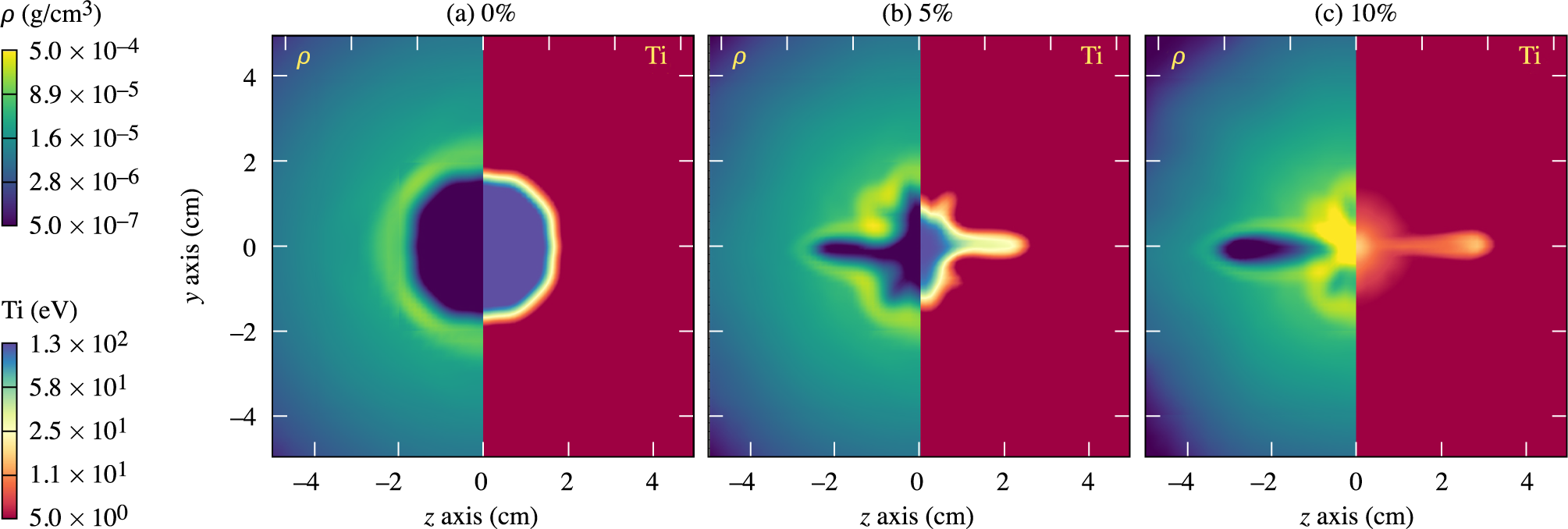}
  \caption{2D slices of mass density in g cm\textsuperscript{-3} (left halves) and ion temperature in eV (right halves) for the 0\%, 5\%, and 10\% perturbation cases at roughly 4.0 $\mu$s.
  This is when the 0\% case reaches $\mathrm{CR} = 20$, which is close to stagnation.}
  \label{fig:dens_tion_stag}
\end{figure}

For these experiment-scale simulations, the minimum fuel radius that can be resolved with 10 computational cells is 1.25 cm, or a convergence ratio (CR) of 20.
The target plasma in the simulations may continue to compress to higher CR, but results become less reliable due to resolution constraints.
Also, higher CRs are not expected to be attainable experimentally, so we consider $\mathrm{CR} = 20$ to be close to stagnation.
Fig.~\ref{fig:dens_tion_stag} shows the mass density (g cm\textsuperscript{-3}) and ion temperature (eV) for all perturbation cases at 4.0 $\mu$s, which is when the 0\% perturbation case reaches $\mathrm{CR} = 20$.
We see that the 0\% case is still mostly spherical at this late time, while the perturbed cases have become distorted.
In the 5\% case, the target plasma is still present but a significant portion has become mixed with the colder liner material resulting in final target ion temperatures $2-3\times$ below the 0\% case.
In both the 0\% and 5\% cases, ion temperatures before stagnation reached peak values ranging from 150-200 eV, but the target remains hotter for longer in the 0\% case.
The target is found to be completely disrupted in the 10\% case as liner plasma has reached the origin.
Based on these results, the 5\% perturbation level is a reasonable upper limit on liner non-homogeneities for maintaining a viable target plasma.
The uniformity of the liner will depend on the jet merging process and physical effects therein, some of which were discussed in Section~\ref{sec:liner}.

Fusion will not occur under the plasma conditions attained in these experiment-scale simulations, but that was expected.
The current instance of the PLX platform is primarily for testing, optimizing, and exploring relevant physics.
Our target compression simulation results show the importance of liner uniformity and that ion temperatures of $\sim$ 200 eV should be possible. Higher temperatures may be within reach with further optimization.
To be a viable option for fusion energy production, the PLX would need to be scaled to a larger system.

\subsection{\label{subsec:flash_reactor}Reactor-scale \emph{FLASH} Simulation}
Here we present a 3D \emph{FLASH} target compression simulation using idealized initial conditions, which are summarized in Table~\ref{tab:3dparams}.
The chosen parameters represent theoretically attainable conditions for a larger, reactor-scale PLX design that employs many more plasma guns for both target and liner formation, hence the much higher densities as compared to the previous experiment-scale simulations.
Furthermore, the reactor-scale simulation assumes that some perfectly-spherical compression has already taken place; the initial target radius is only 4 cm, and the initial liner thickness is 1 cm.
This was done, in part, to accurately model the problem with higher resolution (a minimum cell width of 0.03125 cm) and track the compression down to smaller radii, or higher convergence ratios.
If we consider 25 cm to still be the ``true'' initial target radius, then this simulation is starting at $\mathrm{CR} = 6.25$.
We should also note that the target plasma in this simulation is 50/50 deuterium-tritium (DT) as opposed to pure hydrogen, and it is initialized with a higher temperature of 108 eV.

The initial magnetic field in the target is randomized with the same method as described in the previous subsection.
We purposely aimed for a regime in which $\langle \chi_e \rangle$ and $\langle \beta \rangle$ were both close to 10, which is similar to the experiment-scale regime.
Due to the increased density and temperature, the resulting volume-averaged field strength required to reach this regime was roughly 63 kG.

\begin{figure}
  \includegraphics[width=0.6\linewidth]{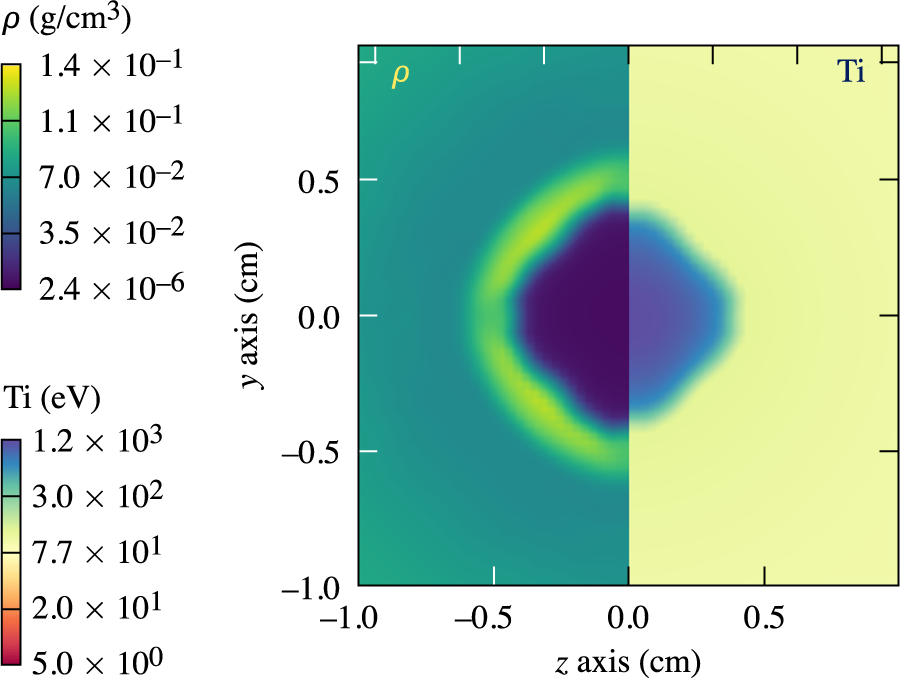}
  \caption{A 2D slice of mass density in g cm\textsuperscript{-3} (left half) and ion temperature in eV (right half) for the 3D reactor-scale \emph{FLASH} simulation at roughly 320 ns.
  This is when the simulation reaches $\mathrm{CR} = 80$, which is close to stagnation.
  No initial liner perturbations were imposed for this simulation.}
  \label{fig:reactor_stag}
\end{figure}

As with the experiment-scale simulations of the previous section, the results of the reactor-scale simulation are reliably accurate as long as the target radius is resolved with at least 10 computational cells.
This means we can resolve a minimum fuel radius of 0.3125 cm, which corresponds to an overall $\mathrm{CR} = 80$.
We consider this CR to be close to stagnation, and it is reached at roughly 320 ns.
Fig.~\ref{fig:reactor_stag} shows the mass density (g cm\textsuperscript{-3}) and ion temperature (eV) at this time, and we observe that the target plasma has reached fusion-relevant conditions.
Peak ion temperatures in the target have exceeded 1 keV, and the volume-averaged ion number density has increased by nearly a factor of 500.
It is also worth noting that the implosion has remained roughly spherical to this high CR, which is perhaps not too surprising since the liner was not perturbed.

Like in the experiment-scale simulations, the target plasma in the reactor-scale simulation lost some magnetic field due to resistive diffusion, but the effect was less pronounced.
Despite starting with a lower electron Hall parameter, as compared to the experiment-scale case, $\langle \chi_e \rangle$ does not go below unity until about 247 ns, which corresponds to a $\mathrm{CR} \approx 19.4$.
This is much later in the compression compared to the experiment-scale 0\% perturbation case, which reached this condition at a CR slightly greater than 1.0.
This is because 1) the reactor-scale liner is moving faster, giving less time for B-field to be lost, and 2) the reactor-scale target is hotter, which results in lower magnetic resistivity.
This supports the idea, presented in the previous section, that some magnetic field losses can be mitigated by increasing the target preheat during target formation.
Using a faster liner may also benefit the magnetic field, but too much interpenetration of the liner jets may not be desirable for the PLX, as mentioned in Section~\ref{subsec:osiris_liner}.

Overall, the 3D reactor-scale simulation results are encouraging, as they show that theoretically-achievable initial conditions with a larger platform can lead to fusion-relevant densities and temperatures.

\section{\label{sec:conc}Conclusions}
In this work, we have presented a comprehensive overview of the PLX platform using three computational codes (\emph{FLASH}, \emph{OSIRIS}, and \emph{HELIOS}) to simulate the three phases of PLX: target formation, liner formation, and target compression.
Each phase has its own unique set of challenges and dominant physical processes.

\emph{FLASH} was used to simulate 2D target formation from two colliding magnetized plasma jets, although the actual experiment employs 4-6 jets.
Jet dynamics were briefly discussed, but our primary focus was on the resulting target conditions.
With jet propagation conditions within experimental ranges, the hydrogen target plasma reached peak ion temperatures of 40 eV and a volume-averaged electron Hall parameter above unity.
The goal of forming a preheated, magnetized target was achieved, and we expect results to further improve when more than two target-forming jets are used.

Using both \emph{FLASH} and \emph{OSIRIS} in 1D, we studied the effects of initial jet velocity on the liner formation process.
The \emph{OSIRIS} results showed that the jets were most collisional at the lowest velocity ($1.3\times 10^6$ cm s$^{-1}$) and collisionless at the highest velocity ($5\times 10^6$ cm s$^{-1}$), thus the \emph{FLASH} results only matched the OSIRIS results for low-velocity cases.
An intermediate, quasi-collisional regime (the $2.5\times 10^6$ cm s$^{-1}$ case) may be desirable, as some interpenetration of the jets led to a smooth density profile and may help mitigate hydrodynamic instabilities.
We also studied the effects of radiation on liner formation using \emph{FLASH}, since \emph{OSIRIS} does not include radiation transport. The simulations showed that radiative losses lead to a liner material that is colder and denser (i.e., more collisional), suggesting that \emph{FLASH} may actually compare well to experimental results of moderate velocities ($2.5\times 10^6$ cm s$^{-1}$).
However, this also suggests that the velocity that results in the desirable quasi-collisional regime may be higher than $2.5\times 10^6$ cm s$^{-1}$.

We explored target compression in 1D with \emph{FLASH}, \emph{OSIRIS}, and \emph{HELIOS}.
All three codes showed similar implosion dynamics with peak compression near 4.5 $\mu$s and fuel temperatures around 150 eV.
Fuel temperatures in the \emph{OSIRIS} simulation were overall slightly lower due to kinetic effects (i.e., mixing of the liner and target materials).
This suggests that the MHD codes may underestimate the level of mixing depending on the relevant plasma conditions when target compression begins, but it was not an effect that dramatically changed the outcomes.

Finally, we conducted high-fidelity 3D target compression simulations with \emph{FLASH} to study the effects of liner perturbations and determined the achievable fuel conditions.
The experiment-scale simulation results were qualitatively similar to their 1D counterparts, with stagnation times near 4.0 $\mu$s and peak ion temperatures in the 150-200 eV range.
Using realistically-positioned liner perturbations, we were able to quantify the target's tolerance to liner non-uniformities.
Results point to a $\sim$ 5\% liner perturbation level (on density and velocity) as a reasonable upper limit for maintaining a mostly spherical implosion and achieving desirable plasma conditions.
An additional, idealized 3D target compression simulation highlighted what may be possible on a larger PLX system:
With higher densities, a slightly higher liner velocity, and higher initial target temperature, this reactor-scale simulation reached fusion-relevant conditions with temperatures above 1 keV.
All of our 3D target compression cases showed evidence of magnetic field losses due to resistive diffusion, which could potentially be mitigated in future PLX experiments.

Future experimental work on the PLX will include further exploration of the parameter space to optimize initial conditions and overall performance.
Our simulation results throughout this work have emphasized the key physics at play, providing guidance for that optimization process.
We have also helped make the case for a larger PLX system, which our work shows may be a theoretically viable path to fusion.
There is still plenty of additional simulation work that can be done, as the PJMIF is such a physics-rich concept.
From a fusion energy production and engineering perspective, the PLX is an attractive design and one that merits further investigation.

\begin{acknowledgments}
This material is based on work supported by the U.S. Department of Energy (DOE) National Nuclear Security Administration (NNSA) under awards DE-NA0004144 and DE-NA0004147, and under subcontracts no. 630138 and C4574 with Los Alamos National Laboratory. We acknowledge support from the U.S. DOE Advanced Research Projects Agency-Energy (ARPA-E) under Award Number DE-AR0001272.
This research used resources of the National Energy Research Scientific Computing Center, a DOE Office of Science User Facility supported by the Office of Science of the U.S. Department of Energy under Contract No. DE-AC02-05CH11231 using NERSC award FES-ERCAP0028680.
The authors acknowledge support from the High-Performance Computing group of the Laboratory for Laser Energetics and the Center for Integrated Research Computing at the University of Rochester.
The software used in this work was developed in part by the U.S. DOE NNSA- and U.S. DOE Office of Science-supported Flash Center for Computational Science at the University of Chicago and the University of Rochester.
This paper describes objective technical results and analysis.
Any subjective views or opinions that might be expressed in the paper do not necessarily represent the views of the U.S. DOE or the United States Government.
\end{acknowledgments}

\bibliography{mylibrary}

\end{document}